\documentstyle[aps,pre,epsfig,times]{revtex}
\begin{document}
\title{Detecting Long-range Correlations with Detrended Fluctuation Analysis}

\author{Jan W. Kantelhardt$^1$, Eva Koscielny-Bunde$^1$, 
Henio H. A. Rego$^1$ [1], Shlomo Havlin$^2$, and Armin Bunde$^1$}
\address{$^1$ Institut f\"ur Theoretische Physik,
Justus-Liebig-Universit\"at Giessen, 
Heinrich-Buff-Ring 16, D-35392 Giessen, Germany}
\address{$^2$ Department of Physics and 
Gonda-Goldschmied-Center for Medical Diagnosis, 
Bar-Ilan University, Ramat-Gan 52900, Israel}
\maketitle

\begin{abstract}
We examine the Detrended Fluctuation Analysis (DFA), which is a 
well-established method for the detection of long-range correlations 
in time series.  We show that deviations from scaling that appear at 
small time scales become stronger in higher orders of DFA, and suggest 
a modified DFA method to remove them.  The improvement is necessary 
especially for short records that are affected by non-stationarities.  
Furthermore, we describe how crossovers in the correlation behavior 
can be detected reliably and determined quantitatively and show how 
several types of trends in the data affect the different orders of DFA.
\end{abstract}

\section{Introduction}

In recent years the Detrended Fluctuation Analysis (DFA) invented 
by Peng et al.\ \cite{peng94} has been established as an important tool 
for the detection of long-range (auto-) correlations in time series
with non-stationarities.  It has successfully been applied to such 
diverse fields of interest as DNA \cite{buldy95,dna,dna1}, heart rate 
dynamics \cite{herz,peng95,herz1,viswan97,peng98,herz2,herz3,herz4,%
herz5,herz6}, neuron spiking \cite{neuron,neuron1}, human gait 
\cite{peng98,gait}, long-time weather records \cite{koscielny98,%
koscielny98b,talkner00}, cloud structure \cite{cloud,cloud1}, economical 
time series \cite{economics,economics1,economics2,economics3}, and even 
solid state physics \cite{fest,fest1}.  While the spectral analysis 
(Fourier transform) and the recently developed wavelet transform modulus 
maxima (WTMM) method \cite{wtmm,wtmm1,wtmm2} analyze the time series 
directly, the DFA is based on random walk theory \cite{randomwalk,randomwalk1}, 
similar to the Hurst rescaled-range analysis \cite{hurst65} (see also 
\cite{feder88}) and similar to another method based on wavelet transform 
used e.~g. in \cite{koscielny98,koscielny98b}.  Since the time series are 
summed in the methods based on the random walk theory (including the
DFA), the noise level due to imperfect measurements in most records
is reduced.  The fano factor \cite{fano} and the allan factor \cite{allan}, 
see also \cite{viswan97}, have been employed in a similar context, 
but these methods do not remove trends in the data.  

For the reliable detection of long-range correlations, it is 
essential to distinguish trends from the long-range fluctuations 
intrinsic in the data.  Trends are caused by external effects -- e.~g. 
the greenhouse warming and seasonal variations for temperature records 
-- and they are usually supposed to have a smooth and monotonous or slowly 
oscillating behavior.  Strong trends in the data can lead to a false detection 
of long-range correlations if only one (non-detrending) method is used or if 
the results are not carefully interpreted.  It is the advantage of the DFA 
that it can systematically eliminate trends of different order (like the 
method based on wavelet transform that has been applied e.~g. in 
\cite{koscielny98,koscielny98b}).  This way we can gain insight into the 
scaling behavior of the natural variability as well as into the trends in the 
considered time series.

In this paper, we study systematically different orders of the DFA technique, 
that allow to eliminate different orders of trends.  The paper is organized 
as follows:  In Section 2 the method is described.  In Section 3 we suggest 
a straightforward extension of the DFA that eliminates DFA specific deviations
from scaling at small time scales.  We describe how crossovers in the observed 
long-range correlation behavior can be detected and detail how the crossover 
time can be determined reliably.  Finally, we show how several types of trends 
in the data affect the different orders of DFA.  We summarize the results in 
the forth section of the paper.

\section{Long-Range Correlations and the Detrended Fluctuation Analysis (DFA)}

We consider a record $(x_i)$ of $i=1,\ldots,N$ equidistant measurements.  In 
most applications, the index $i$ will correspond to the time of the measurements.
We are interested in the correlation of the values $x_i$ and $x_{i+s}$ 
for different time lags, i.~e. correlations over different time scales $s$.  
In order to get rid of a constant offset in the data, the mean 
$\langle x \rangle = {1 \over N} \sum_{i=1}^{N} x_i$ is usually subtracted, 
$\bar{x}_i \equiv x_i - \langle x \rangle$.  Quantitatively, correlations 
between $x$-values separated by $s$ steps are defined by the (auto-) correlation 
function
\begin{equation} C(s) = \big\langle \bar{x}_i \, \bar{x}_{i+s}\big\rangle 
= {1 \over N-s} \sum_{i=1}^{N-s} \bar{x}_i \, \bar{x}_{i+s}.
\label{autocorr}\end{equation}
If the $x_i$ are uncorrelated, $C(s)$ is zero for $s>0$.  Short-range 
correlations of the $x_i$ are described by $C(s)$ declining exponentially,
$C(t) \sim \exp (-s/s_\times)$ with a decay time $s_\times$.  
For so-called long-range correlations $C(s)$ declines as a power-law
\begin{equation} C(s) \propto s^{- \gamma} 
\label{gamma}\end{equation}
with an exponent $0 < \gamma < 1$.  A direct calculation of $C(s)$ is usually 
not appropriate due to noise superimposed on the collected data $x_i$ and due 
to underlying trends of unknown origin.  For example, the average $\langle x 
\rangle$ might be different for the first and the second half of the record,
if the data are strongly long-range correlated.  This makes the definition of 
$C(s)$ problematic.  Thus, we have to determine the correlation exponent 
$\gamma$ indirectly.

Often experimental data are affected by non-stationarities.  Such trends have 
to be well distinguished from the intrinsic fluctuations of the system in order 
to find the correct scaling behavior of the fluctuations.  This task is not 
easy, since e.~g. subtracting some kind of moving average with a certain
bin width $\sigma$ would artificially introduce the time scale $\sigma$ into the 
data, thus destroying a possible scaling over a wider range of time scales.  
Hurst rescaled-range analysis \cite{hurst65} and other non-detrending methods 
\cite{fano,allan} work well if the records are long and do not involve trends.  
But if trends are present in the data, they might give wrong results.  Very often 
we do not know the reasons for underlying trends in collected data and -- even 
worse -- we do not know the scales of the underlying trends.  Detrended 
fluctuation analysis (DFA) is a well-established method for determining the 
scaling behavior of noisy data in the presence of trends without knowing their 
origin and shape \cite{peng94,buldy95,peng95}.  

The DFA procedure consists of four steps.  In the first step, we determine 
the profile
\begin{equation} Y(i) = \sum_{k=1}^i x_k - \langle x \rangle
\label{profile} \end{equation}
of the record $(x_i)$ of length $N$.  The subtraction of the mean $\langle x 
\rangle$ is not compulsory, since it would be eliminated by the later detrending 
in the third step anyway.

\begin{figure} \begin{center}
\epsfig{file=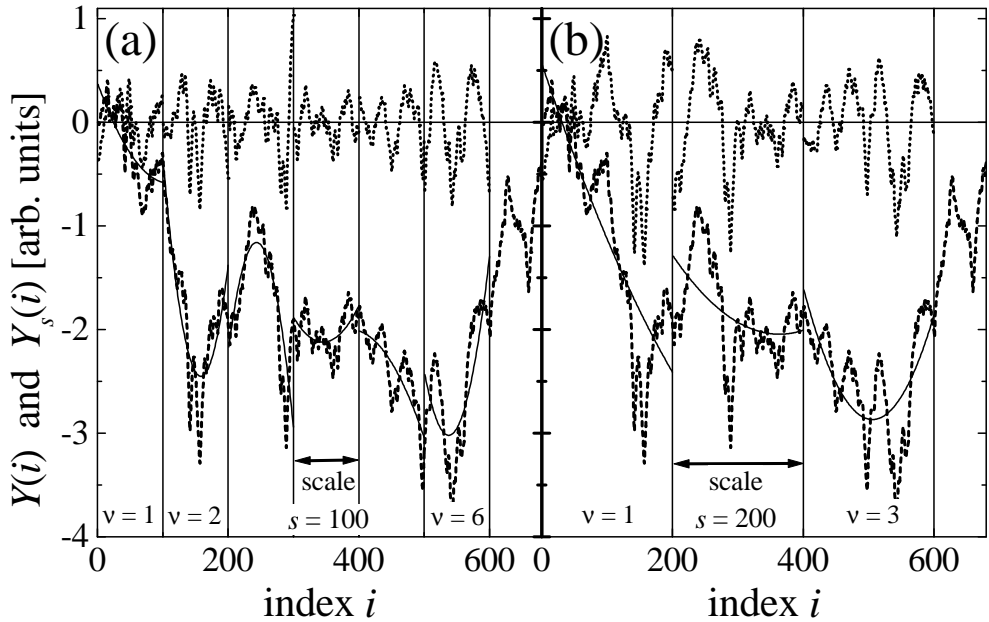,width=13cm} 
\end{center}
\caption{Illustration of the detrending procedure in the Detrended Fluctuation 
Analysis (DFA).  For two segment lengths (time scales) $s = 100$ (a) and
$200$ (b), the profiles $Y(i)$ (dashed lines; defined in
Eq.~(\protect\ref{profile})), least square quadratic fits to the profiles (solid
lines), and the detrended profiles $Y_s(i)$ (dotted lines) are shown versus
the index $i$.}
\label{dfa} \end{figure}

In the second step we cut the profile $Y(i)$ into $N_s \equiv [N/s]$ 
non-overlapping segments of equal length $s$ (see Fig.~\ref{dfa}).  
Since the record length $N$ need not be a multiple of the considered time 
scale $s$, a short part at the end of the profile will remain in most cases.  
In order not to disregard this part of the record, the same procedure is 
repeated starting from the other end of the record.  Thus, $2 N_s$ segments 
are obtained altogether.

In the third step we calculate the local trend for each segment $\nu$ by a 
least-square fit of the data.  Then we define the detrended time series 
for segment duration $s$, denoted by $Y_s(i)$, as the difference between 
the original time series and the fits,
\begin{equation} Y_s(i) = Y(i) - p_{\nu}(i),
\label{detrprofile} \end{equation}
where $p_{\nu}(i)$ is the fitting polynomial in the $\nu$th segment.  Figure 
\ref{dfa} illustrates this procedure for $s = 100$ and 200.  In the example 
quadratic polynomials are used in the fitting procedure, which is characteristic 
of quadratic DFA (DFA2).  Linear, cubic, or higher order polynomials can also 
be used in the fitting procedure (DFA1, DFA3, and higher order DFA).  
Since the detrending of the time series is done by the subtraction of the
fits from the profile, these methods differ in their capability of eliminating
trends in the data.  In $n$th order DFA, trends of order $n$ in the profile and
of order $n - 1$ in the original record are eliminated.  Thus a comparison of
the results for different orders of DFA allows to estimate the strength of the
trends in the time series, as will be shown in Section \ref{dettrend}.  

In the fourth step, we calculate -- for each of the $2 N_s$ segments -- the 
variance 
\begin{equation} F^2_s(\nu) = \langle Y^2_s(i) \rangle = {1 \over s} 
\sum_{i=1}^{s} Y^2_s[(\nu-1) s + i] \label{fsdef} \end{equation}
of the detrended time series $Y_s(i)$ by averaging over all data points $i$ in 
the $\nu$th segment.  At last we average over all segments and take the square 
root to obtain the DFA fluctuation function
\begin{equation} F(s) = \left[ {1 \over 2 N_s} \sum_{\nu=1}^{2 N_s} 
F^2_s(\nu) \right]^{1/2} \label{fdef} \end{equation}
For different detrending orders $n$ we obtain different fluctuation functions 
$F(s)$, which we then denote by $F^{(n)}(s)$.  By construction, $F^{(n)}(s)$
is only defined for $s \ge n+2$.  We are interested in the $s$-dependence of
$F^{(n)}(s)$.  It is apparent, that the variance will increase with increasing
duration $s$ of the segments.  If the data $(x_i)$ are long-range power-law 
correlated [see Eq.~(\ref{gamma})], the fluctuation functions $F^{(n)}(s)$ 
increase by a power-law
\begin{equation} F^{(n)}(s) \propto s^\alpha \label{alpha} \end{equation}
for large $s$ values, where the fluctuation exponent $\alpha$ is related to 
the correlations exponent $\gamma$.  A direct derivation of Eq.~(\ref{alpha}) 
for large $s$ is given by \cite{taqqu95} in the appendix.  For data without 
trends and zero offset, $\bar{x}_i = x_i$ and $Y_s(i) = Y(i)$ for $i \le s$.  
Then the mean-square displacement in each segment $n$ can be calculated 
[with Eqs.~(\ref{profile}) and (\ref{autocorr})]:
\begin{equation} \langle Y^2(i) \rangle =  \left\langle \sum_{k=1}^i  x_k^2 
\right\rangle + \left\langle \sum_{k \ne j}^{j,k \le i} x_j \, x_k \right\rangle 
= i \langle x^2 \rangle + \sum_{k \ne j}^{j,k \le i} C(\vert k - j \vert) 
= i \langle x^2 \rangle + 2 \sum_{k=1}^{i-1} (i-k) C(k).
\label{calc1}\end{equation}
For large $i$ the second term can be approximated [with Eq.~(\ref{gamma})]:  
\begin{equation} \sum_{k=1}^{i-1} C(k) 
\sim \sum_{k=1}^{i} k^{-\gamma} \sim \int_1^{i} k^{-\gamma} dk 
\sim i^{1-\gamma} \quad {\rm and} \quad \sum_{k=1}^{i-1} k C(k) 
\sim i^{2-\gamma}. \end{equation}
If the data are long-range power-law correlated with $0 < \gamma < 1$, this term 
will dominate for large $i$, giving
\begin{equation} \langle Y^2(i) \rangle \sim i^{2-\gamma}. 
\label{Ygamma} \end{equation}
Thus, the mean-square displacement $\langle Y^2(i) \rangle$ of the profile 
increases faster than linearly in $i$, which corresponds to superdiffusion.
A similar approximation for $F^{(n)}(s)$ using Eq.~(\ref{Ygamma}) finally 
leads to
\begin{equation} F^{(n)}(s) \sim s^{1-\gamma/2} \label{Fgamma} \end{equation}
for large time scales $s$ (see the appendix of \cite{taqqu95} for an exact 
derivation for DFA1).  Thus, comparing Eqs.~(\ref{alpha}) and (\ref{Fgamma}), 
we find 
\begin{equation} \alpha = 1 - \gamma/2 \quad {\rm for} 
\quad 0 < \gamma < 1. \label{algam} \end{equation}
If the data are uncorrelated or short-range correlated [$C(s)$ decays 
exponentially or $\gamma > 1$ in Eq.~(\ref{gamma})], the first term in 
Eq.~(\ref{calc1}) will dominate for large $i$, and we find $\langle Y^2(i) 
\rangle \sim i$ (corresponding to regular diffusion) and hence 
$F^{(n)}(s) \sim s^{1/2}$.  Thus, the fluctuation exponent $\alpha = 1/2$ 
indicates the absence of long-range correlations.

\begin{figure} \begin{center}
\epsfig{file=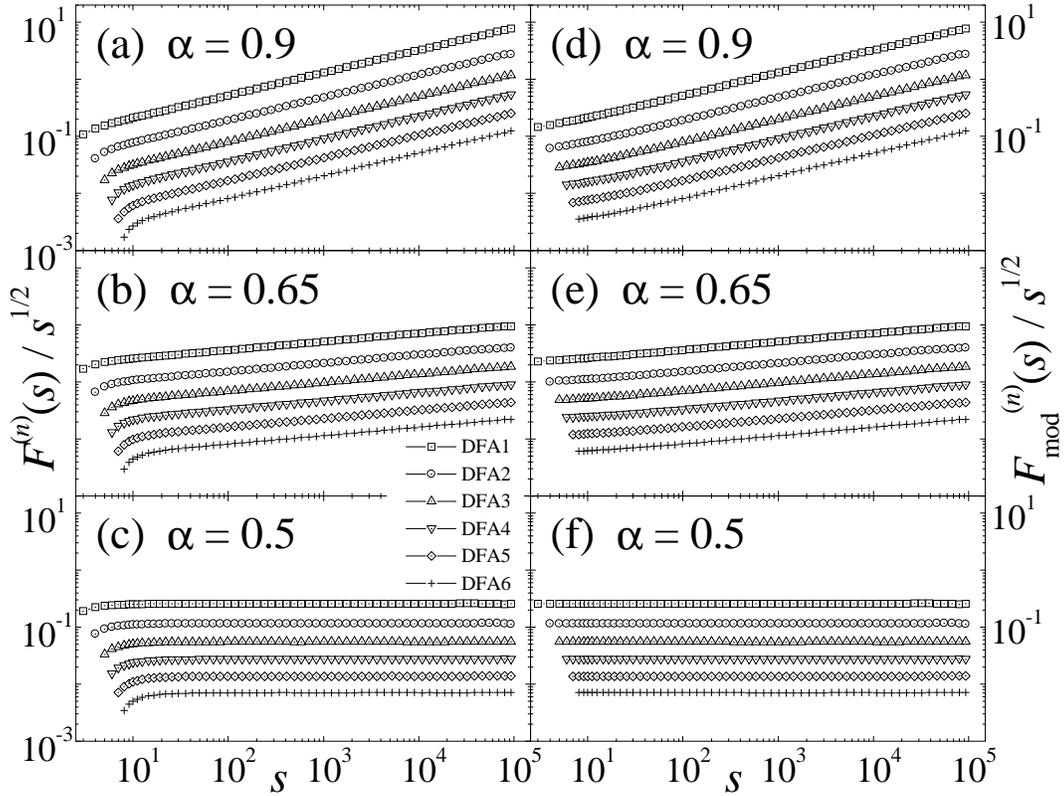,width=14cm} 
\end{center}
\caption{Detrended Fluctuation Analysis (DFA) of long-range correlated 
data with (a,d) $\gamma = 0.2$ ($\alpha = 0.9$), (b,e) $\gamma = 0.7$ 
($\alpha = 0.65$), and uncorrelated data (c,f) with $\alpha = 0.5$.  
The scaled common DFA fluctuation functions $F^{(n)}(s) /s^{1/2}$ 
are plotted versus the time scale $s$ in (a,b,c), while the corresponding 
modified DFA fluctuation functions $F^{(n)}_{\rm mod}(s)$ are shown in 
(d,e,f).  The symbols correspond to the different detrending orders $n$, 
DFA1 ($\Box$), DFA2 ({\Large$\circ$}), DFA3 ($\bigtriangleup$), 
DFA4 ($\bigtriangledown$), DFA5 ($\Diamond$), and DFA6 ($+$).  
The artificial long-range correlated data have been generated by the 
Fourier transform method, see e.~g. \protect\cite{feder88,makse96}. 
The deviations from scaling occurring in $F(s)$ for small time scales $s$ in 
parts (a,b,c) are drastically reduced by dividing by the correction 
function obtained from shuffled data in the parts (d,e,f).  The results have
been obtained by averaging over 100 artificial series of length $N=200,000$
for each part.}
\label{res1} \end{figure}

Practically, we can plot $F^{(n)}(s)$ as a function of $s$ on double logarithmic 
scales to measure $\alpha$ by a linear fit.  For uncorrelated or short-range 
correlated data, we expect $\alpha = 0.5$, while $\alpha > 0.5$ 
indicates long-range correlations.  In this case we can determine the 
correlation exponent $\gamma$ by measuring the fluctuation exponent 
$\alpha$.  Figure~\ref{res1} shows two examples for the application 
of the DFA method to long-range correlated data [Fig.~\ref{res1}(a,b)]. 
An example for uncorrelated artificial data [Fig.~\ref{res1}(c)] is also 
shown, confirming $\alpha = 1/2$.

\section{Results}
\subsection{The Correction Function and an Modified Version of the DFA}

Figures \ref{res1}(a,b,c) show that small deviations from the scaling law 
Eq.~(\ref{alpha}), i.~e. deviations from a straight line in the log-log plot, 
occur for small scales $s$.  These deviations are intrinsic to the usual DFA 
method, since the scaling behavior is only approached asymptotically.  The 
deviations limit the capability of DFA to determine the correct correlation 
behavior in very short records and in the regime of small $s$.  DFA6, e.~g., 
is only defined for $s \ge 8$, and significant deviations from the scaling 
law (\ref{alpha}) occur even up to $s \approx 30$.  They will lead to an 
over-estimation of the fluctuation exponent $\alpha$, if the regime of small 
$s$ is used in a fitting procedure.  Previously, an attempt has been made to 
improve the scaling by modifying the prefactor in the definition of $F(s)$ 
in Eq.~(\ref{fdef}) \cite{buldy95}, but it is valid for DFA1 only and it still 
contains some approximations.  Here, we suggest a different approach, which 
can also be applied for higher order DFAs.  

\begin{figure} \begin{center}
\epsfig{file=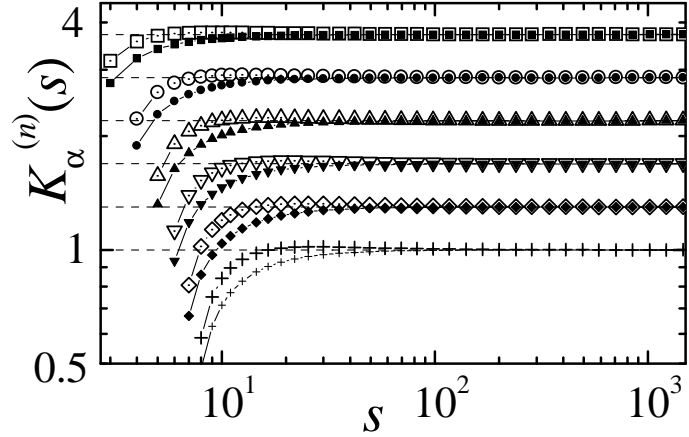,width=9cm} 
\end{center}
\caption{The DFA correction functions $K^{(n)}_\alpha(s)$ for uncorrelated 
data ($\alpha = 1/2$, small filled symbols) and for long-range correlated data 
($\alpha = 0.9$, large open symbols) are shown versus $s$.  The 
symbols correspond to the different detrending orders of DFA (see 
Fig.~\protect\ref{res1} for the definition) and the functions are shifted 
by multiples of 1.3.  The dashed lines indicate the asymptotic behavior.  
$K^{(n)}_\alpha(s) \to 1$.  In the numerical procedure, $K^{(n)}_\alpha(s)$ 
has been obtained by averaging over 100 artificial series of length 
$N=200,000$.  The long-range correlated series have been generated by the 
method of Fourier transform, see e.~g. \protect\cite{feder88,makse96}. }
\label{res2} \end{figure}

For long artificial series of uncorrelated ($\alpha = 1/2$) and long-range 
correlated data (with $\alpha = 0.9$) we have determined the deviations of 
$F^{(n)}(s)$ from the expected scaling behavior Eq.~(\ref{alpha}).  
Figure \ref{res2} shows the results for the correction function
\begin{equation} K^{(n)}_\alpha(s) = {\langle [F^{(n)}(s)]^2 \rangle^{1/2} 
\, s'^\alpha \over \langle [F^{(n)}(s')]^2 \rangle^{1/2} \, s^\alpha } 
\quad {\rm (for} \, s' \gg 1),
\label{defk}\end{equation}
where, again, $n$ denotes the DFA detrending order, and $\langle \ldots \rangle$ 
denotes the average over different configurations.  Practically, $s'$ has to be
large ($s' > 50$), but it must remain significantly smaller than the record 
length $N$; $s' \approx N/20$ seems to be a reasonable number.  If we divide 
the DFA fluctuation functions $F(s)$ by the corresponding correction function 
$K^{(n)}_\alpha(s)$, the deviations from scaling for small $s$ are eliminated.  
The crucial point is that the correction function $K^{(n)}_\alpha(s)$ 
depends only weakly on $\alpha$ (see Fig.~\ref{res2}).  Therefore, practically, 
the correction function for uncorrelated data, $K^{(n)}_{1/2}(s)$, can be 
used in all cases.  $K^{(n)}_{1/2}(s)$ can be obtained most easily by 
analyzing the corresponding shuffled data, where all long-range correlations 
have been destroyed by randomly shuffling the record of measurements.  

Thus, in order to improve the scaling of the DFA fluctuations on short scales 
$s$, we suggest to divide $F^{(n)}(s)$ by $K^{(n)}_{1/2}(s)$ to obtain 
the {\it modified} fluctuation function
\begin{equation} F^{(n)}_{\rm mod}(s) = {F^{(n)}(s) \over K^{(n)}_{1/2}(s)}
= F^{(n)}(s) {\langle [F^{(n)}_{\rm shuff}(s')]^2 \rangle^{1/2} \, s^{1/2} 
\over \langle [F^{(n)}_{\rm shuff}(s)]^2 \rangle^{1/2} \, s'^{1/2} } 
\quad {\rm (for} \, s' \gg 1),
\label{fmod}\end{equation}
according to Eq.~(\ref{defk}).  Here, $\langle [F^{(n)}_{\rm shuff}(s)]^2 
\rangle^{1/2}$ denotes the usual DFA fluctuation function [defined in 
Eq.~(\ref{fdef})] averaged over several configurations of shuffled data
taken from the original record $(x_i)$ under consideration, and 
$s' \approx N/20$, again.

In Fig.~\ref{res1} the numerical results for $F^{(n)}_{\rm mod}(s)$ are
compared to those for the common $F^{(n)}(s)$.  The improvement of the
scaling behavior for small $s$ can be seen clearly by comparing the right 
and left parts of the figure.  Only very weak deviations from the expected 
power-law dependence remain for strong long-range correlations [for large 
$\alpha$, see Fig.~\ref{res1}(d)].  The suggested modified
method does only need additional computation time, but the programming 
effort is not significantly higher.  Thus, it can easily be used in most 
applications.  The improvement is very useful especially for short records
or records that have to be split into shorter parts to eliminate problematic 
nonstationarities, since the small $s$ regime can be included in the fitting 
range for the fluctuation exponent $\alpha$.

Note that there is another advantage in dividing by the 
DFA fluctuation function for shuffled data:  If the distribution of the 
$x$-values in the record is very broad, e.~g. similar to a Levy
distribution, and the second moment $\langle x^2 \rangle$ diverges, 
systematic deviations from the expected scaling behavior will occur already 
for uncorrelated data (see e.~g. \cite{levywalk}).  Since these deviation 
are {\it not} eliminated by shuffling the data, they will cancel out in 
$F^{(n)}_{\rm mod}(s)$.  Thus, the modified DFA method indicates the correct 
correlation behavior also in presence of broadly distributed data, where 
the common DFA fails to distinguish long-range correlations from deviations
caused by broad distributions.

\subsection{Determination of Crossovers}
\label{detcross}

Frequently, the correlations of recorded data do not follow the same scaling
law for all time scales $s$, but one or sometimes even more crossovers between
different scaling regimes are observed.  For example, the data might become 
uncorrelated on large time scales $s > s_\times$.  In such a case, it would 
be useful to extract the crossover time $s_\times$ from the data also by
means of DFA.  In order to do this, we have to investigate how crossovers 
in the correlation properties show up in the DFA fluctuation functions with
different orders of detrending.  

Artificial time series with a well-defined crossover at $s_\times$ are 
most easily generated in a modified Fourier transform procedure:  The power 
spectrum $P(f)$ of an uncorrelated random series is multiplied by 
$(f/f_\times)^{-\beta}$ with $\beta = 2 \alpha -1$ for frequencies 
$f > f_\times = 1/s_\times$ only.  The series obtained by inverse Fourier 
transform of this modified power spectrum exhibits power-law correlations on 
time scales $s < s_\times$ only, while the behavior becomes uncorrelated on 
larger time scales $s > s_\times$.  An inverse crossover with long-range 
correlations only for $s > s_\times$ and uncorrelated behavior below 
$s_\times$ is obtained in a similar way, if we multiply the power spectrum 
by $(f/f_\times)^{-\beta}$ for low frequencies $f < f_\times$ only.  
Note that there is an alternative way to generate series with a crossover 
in the correlation behavior:  We can divide the original long-range 
correlated series into segments of length $s_\times$ and shuffle the segments.  
This way, all correlations for $s>s_\times$ are destroyed, but the 
correlations within the segments, i.~e. for $s<s_\times$, are preserved.  
%Alternatively, the data within each segment can be shuffled, destroying the 
%correlations for $t<t_\times$, but leaving them intact for $t>t_\times$.  
With this method the crossover regime turns out to be much broader, so we 
use the Fourier transfrom method in this paper.

\begin{figure} \begin{center}
\epsfig{file=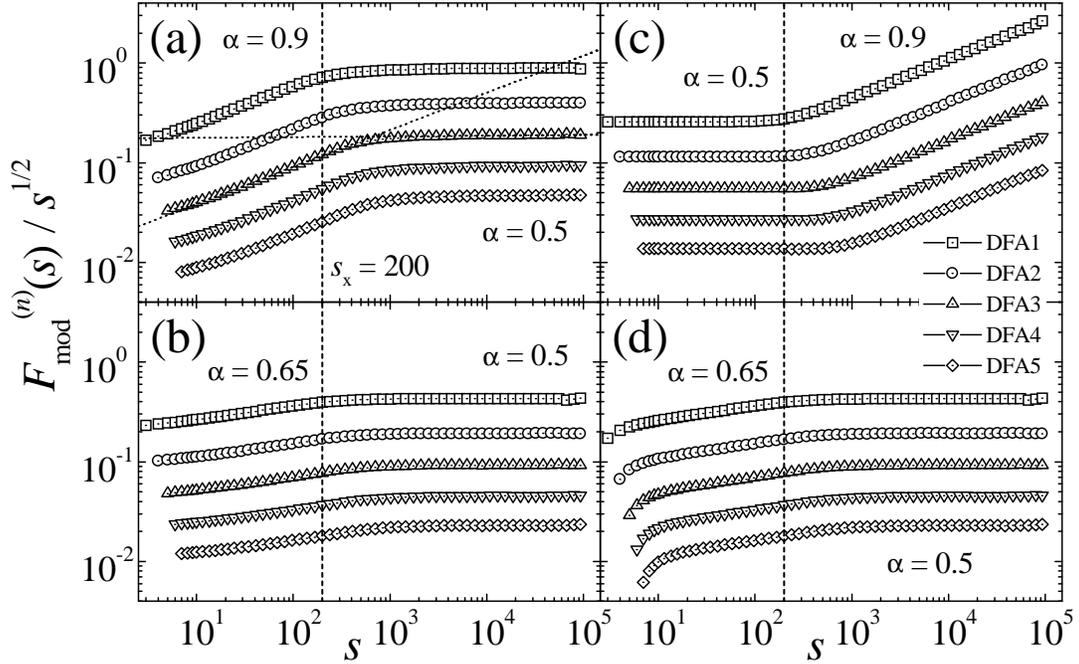,width=14cm} 
\end{center}
\caption{Modified DFA of artificial series with a crossover from long-range 
correlated behavior [(a): $\gamma = 0.2$, $\alpha = 0.9$; (b): $\gamma = 0.7$, 
$\alpha = 0.65$] for $s < s_\times = 200$ to uncorrelated behavior for 
$s > s_\times$.  The scaled modified DFA fluctuation functions 
$F^{(n)}_{\rm mod}(s) /s^{1/2}$ are plotted versus the time scale $s$ for 
DFA1 to DFA5.  In part (c) the results for data with an inverse crossover 
from uncorrelated behavior (for $s < s_\times = 200$) to long-range correlated 
behavior ($\gamma = 0.2$, $\alpha = 0.9$ for $s > s_\times$) are shown.  For 
comparison, part (d) shows the usual DFA fluctuation functions $F(s)$ for the 
artificial series already considered in part (b).  The long-range correlated 
series with crossover have been generated by the modified Fourier transform 
method described in the text, and the results for 100 series of length 
$N=200,000$ have been averaged.  The symbols correspond to the different 
orders of DFA (see Fig.~\protect\ref{res1} for the definition).  The dotted
fits in part (a) illustrate the procedure used to determine the observed 
position of the crossover $s_\times^{(3)}$, while the dashed vertical lines
indicate the real position of the crossover $s_\times = 200$.}
\label{res3} \end{figure}

Figure~\ref{res3} shows the results for the modified DFA fluctuation function 
$F^{(n)}_{\rm mod}(s)$, see Eqs.~(\ref{fdef}) and (\ref{fmod}), for artificial
data with a crossover in the correlation behavior.  The crossover is clearly 
visible in the results, but it occurs at times $s_\times^{(n)}$ that depend on
the detrending order $n$ and that are different from the original $s_\times$ 
used for the generation of the data.  This systematic deviation is most 
significant in the DFA$n$ with higher order detrending.  It occurs independent 
of the values of the fluctuation exponents and independent of the direction of 
the crossover (from small to large exponents or vice versa).  

\begin{figure} \begin{center}
\epsfig{file=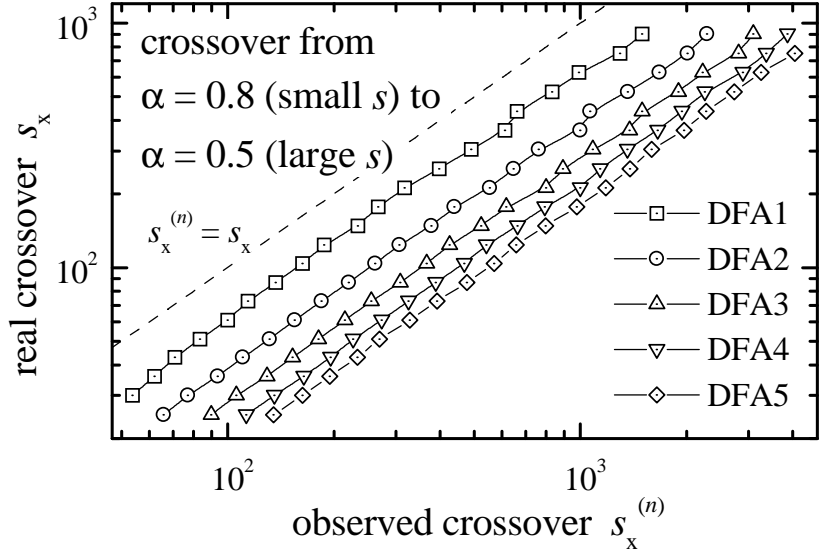,width=11cm} 
\end{center}
\caption{Systematic investigation of the crossover detection with the modified 
DFA1 to DFA5.  For artificial data with a crossover from long-range correlated 
behavior ($\gamma = 0.4$, $\alpha = 0.8$) for $s < s_\times$ to uncorrelated 
behavior for $s > s_\times$ the crossover times $s_\times^{(n)}$ have 
been determined from the intersection of linear fits done on both sides of the 
crossovers [as illustrated in Fig.~\protect\ref{res3}(a)].  The original 
crossover time $s_\times$ is varied and plotted versus $s_\times^{(n)}$ for 
all five types of DFA.  Note that the results also hold for inverse crossovers 
and for other values of $\alpha$, although the fitting procedure will be less 
accurate if $\alpha$ is close to 0.5.  They also hold for the usual DFA for 
$s_\times > 200$, since $K^{(n)}_{1/2}(s) \approx 1$ for large $s$.  The
dashed line corresponds to $s_\times^{(n)} = s_\times$, and its position shows 
that the estimated crossovers are always larger than the real $s_\times$.  The 
long-range correlated series with crossover have been generated by the modified 
Fourier transform method described in the text.  For each point the results for 
200 time series of length $N=100,000$ have been averaged, and the error bars have
approximately the size of the symbols.  The symbols correspond to the different 
orders of DFA (see Fig.~\protect\ref{res1} for the definition). }
\label{res4} \end{figure}

The deviation of the crossovers is systematically investigated in 
Fig.~\ref{res4}, where the position of the original crossover $s_\times$ 
used for the data generation is plotted versus the position of observed 
crossover $s_\times^{(n)}$ for DFA1 to DFA5.  The plot can be used to 
determine the real crossover position $s_\times$ from the $s_\times^{(n)}$ 
estimated with the modified DFA$n$.  If several orders of DFA are used in 
this procedure, several estimates will be obtained which can be checked 
for consistency or used for an error approximation.

\begin{figure} \begin{center}
\epsfig{file=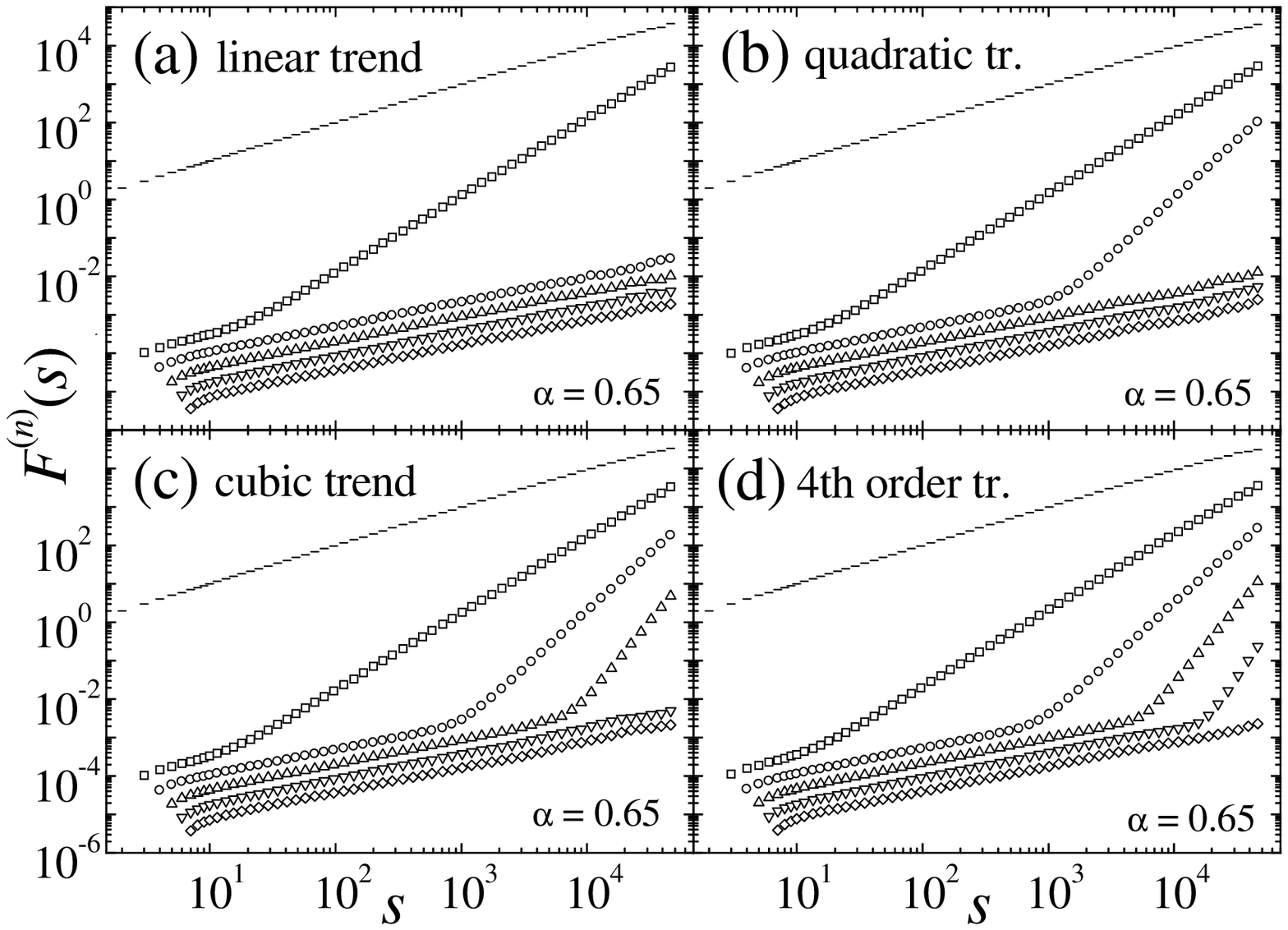,width=14cm} 
\end{center}
\caption{Investigation of the trend elimination and detection capability of 
the conventional Fluctuation Analysis (FA, symbol $-$) and the (not modified) 
DFA1 to DFA5 for trends $A x^p$ with integer powers $p$.  For artificial 
long-range correlated data ($\gamma = 0.7$, $\alpha = 0.65$) with added 
(a) linear, (b) quadratic, (c) cubic, and (d) 4th order trends, the DFA 
fluctuation functions $F^{(n)}(s)$ are plotted versus the time scale $s$ 
($A=10^4$, $p=1,2,3,4$).  The trends are completely eliminated if $n > p$.  
This feature allows to determine the (integer) order $p$ of the trends using 
DFA.  For $n \le p$ the trends lead to an apparent crossover at high $s$ 
values as discussed in the text.  Only one series of length $N=100,000$ has 
been considered for each part of the figure.  The symbols correspond to the 
different orders of DFA (see Fig.~\protect\ref{res1} for the definition). }
\label{res5a} \end{figure}

\subsection{Monotonous Trends} \label{dettrend}

Records from real measurements are often affected by trends, which have to 
be well distinguished from the intrinsic fluctuations of the system as 
discussed in Section 2.  To investigate the effect of trends on the DFA 
fluctuation functions, we have generated artificial series $(x_i')$ with smooth 
monotonous trends by adding polynomials of different power $p$ to the original
record $(x_i)$ generated with the Fourier transform method: 
\begin{equation}
x_i' = x_i + A x^p \qquad {\rm with} \quad x = i/N.
\end{equation}

Figure \ref{res5a} shows the effect of trends of different (integer) order $p$ 
on the FA and DFA fluctuation functions $F^{(n)}(s)$.  For the conventional
Fluctuation Analysis (FA), the variance $F^2_s(\nu)$ for segment $\nu$ of scale 
length $s$ is determined by the mean square deviation of the (non-detrended) 
profile $Y(i)$, i.~e. Eq.~(\ref{fsdef}) is replaced by
\begin{equation}
F^2_s(\nu) = [ Y(\nu s) - Y(\nu s -s+1) ]^2.
\end{equation}

\begin{figure} \begin{center}
\epsfig{file=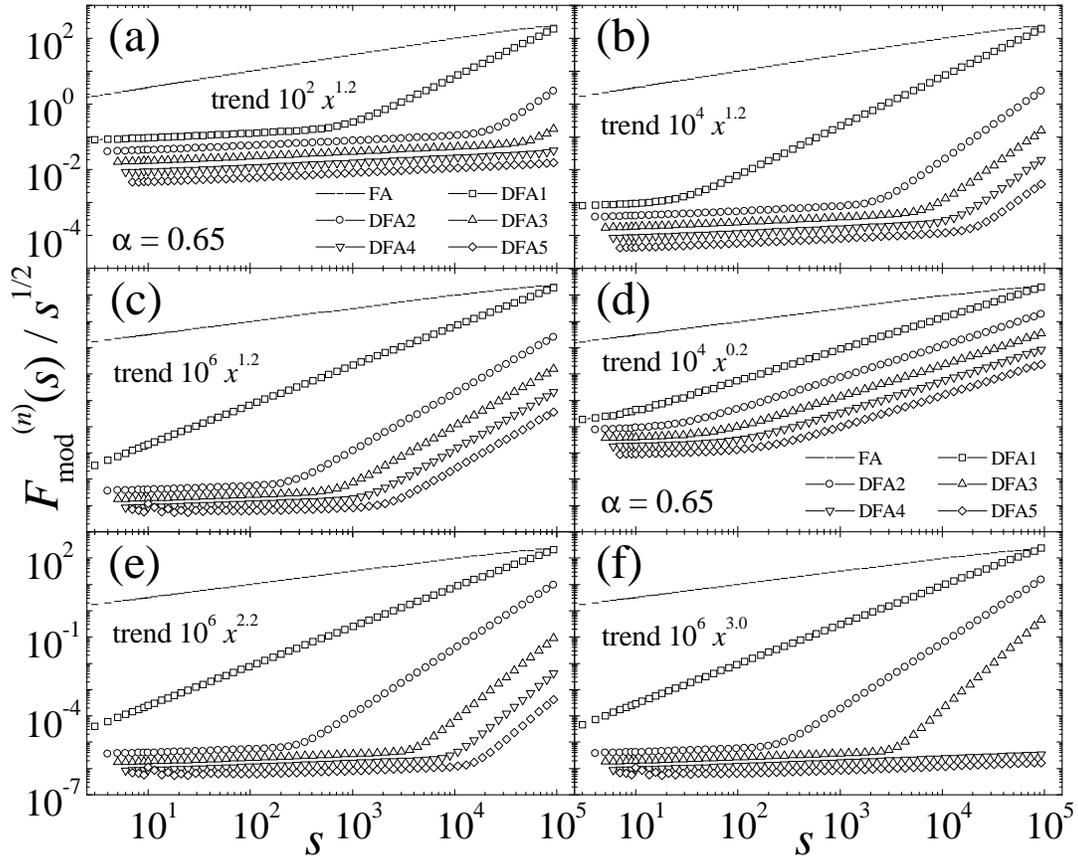,width=14cm} 
\end{center}
\caption{Investigation of the trend elimination and detection capability of 
the modified DFA1 to DFA5 for trends $A x^p$ with non-integer powers $p$.  
For artificial long-range correlated data ($\gamma = 0.7$, $\alpha = 0.65$) with 
added trends (as listed in the figure), the scaled modified DFA fluctuation 
functions $F^{(n)}_{\rm mod}(s) /s^{1/2}$ are plotted versus the time scale $s$.  
As expected, the trends are best eliminated by the higher order DFAs, but an 
apparent crossover is still observed for large time scales $s$.  The position 
of the crossover and the slopes of the curves above the crossover can be used 
to determine the strength $A$ and the exponent $p$ of the trends (see text). 
The results for 100 series of length $N=200,000$ have been averaged.  The symbols 
are the same as in Fig.~\protect\ref{res5a}. }
\label{res5} \end{figure}

This procedure leads to similar results as the DFA, but trends are not 
eliminated.  In Fig.~\ref{res5a} the strong trends completely conceal the 
long-range correlations of the time series.  The slope $\alpha_{\rm trend} = 1$ 
is observed, which is the maximum for the FA, characteristic of strong trends
(or non-stationary time series).  For the DFA, the trends in the data can lead 
to an artificial crossover in the scaling behavior of $F^{(n)}(s)$, i.~e. 
the slope $\alpha$ is increased for large time scales $s$.  The position of 
this artificial crossover depends on the strength $A$ and the power $p$ of 
the trend.  Evidently, no artificial crossover is observed, if the detrending 
order $n$ is larger than $p$.  Thus, the order $p$ of the trends in the 
data can be determined easily by applying the different DFA$n$.  If $p$ 
is larger than $n$, an artificial crossover is observed, and the slope 
$\alpha_{\rm trend}$ in the large $s$ regime strongly depends on $n$.  Thus, 
the artificial crossover can be clearly distinguished from real crossovers in 
the correlation behavior, which would result in identical slopes $\alpha$ and 
rather similar crossover positions for all detrending orders $n$ (see
Fig.~\ref{res3}).  

Figure \ref{res5} shows the same type of results as Figs.~\ref{res5a}, but for 
the modified DFA and for trends $A x^p$ with non-integer power $p$.  Here, 
strong trends are not completely eliminated by the DFA and can dominate the 
behavior of the fluctuation functions on very large time scales $s$ even if
$n > p$.  If the trends are not too weak [as for the DFA5 in Fig.~\ref{res5}(a)] 
or too strong [as for the DFA1 in Fig.~\ref{res5}(c,e,f)] an artificial crossover
occurs.  The positions of the artificial crossover and the 
slopes $\alpha_{\rm trend}^{(n)}$ in the large $s$ regime can be used to 
determine the trend parameters $A$ and $p$.  Two main rules can be deduced 
from the results shown in Fig.~\ref{res5}, and we have confirmed these rules 
also for other parameters $\alpha$, $A$ and $p$:  (i) If a trend-related 
crossover is observed and the detrending order $n$ is sufficiently small 
($n < p+0.5$), the position of the artificial crossover 
depends only on $A$, but not on $p$ or $\alpha$ [compare Figs.~\ref{res5}(c,e,f)].  
This permits to determine the strength $A$ of the trend for real data by 
comparison to the results for artificial data with known trend strength [see 
Figs.~\ref{res5}(a,b,c)].  (ii) If a trend-related crossover is observed, the slope 
$\alpha_{\rm trend}^{(n)}$ for large $s$ (above this crossover) is the minimum 
of $n+1$ and $p+1.5$ [see Fig.~\ref{res5}(c,d,e)].  This permits to determine 
the order $p$ of the trend, if $n$ is chosen sufficiently large.  The 
combination of both rules allows to determine $A$ and $p$ if the trend is strong 
enough and several appropriate DFA orders $n$ are employed.  Note that the 
high order DFAs tend to become numerically unstable on small scales $s$ for 
very strong trends and weak fluctuations [see DFA7 for $s<30$ in 
Fig.~\ref{res5}(c,e,f)].

\begin{figure} \begin{center}
\epsfig{file=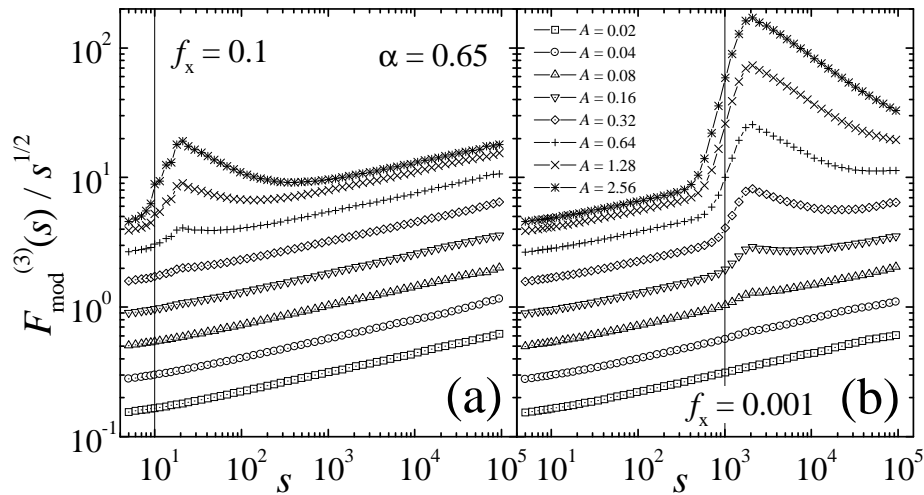,width=12cm} 
\end{center}
\caption{Modified DFA of long-range correlated random data ($\alpha = 0.65$, 
unit variance) with additional oscillatory trends $A \sin(2 \pi i f)$ of 
different frequencies $f$, (a) $f=0.1$ and (b) $f=0.001$, and intensities $A$
ranging from $A=0.02$ to $A=2.56$.  The modified DFA fluctuation functions 
$F^{(3)}_{\rm mod}(s)$ are plotted versus the time scale $s$.  For other
orders of the DFA we get similar results.  Only for strong oscillatory trends 
the expected scaling behavior $F^{(n)}_{\rm mod}(s) \sim s^{0.65}$ is 
disturbed.  This shows that DFA is quite robust against oscillatory trends, 
especially against those with rather high frequencies.  The results for 100 
series of length $N=200,000$ have been averaged for each curve. }
\label{res6} \end{figure}

\subsection{Oscillatory Trends}

Sometimes the trends superimposed on the records of real measurements are 
not monotonous.  A quite common shape of trends is slow oscillatory
behavior.  It can be either rather periodic and regular, e.~g. for the 
seasonal trend in temperature fluctuations, or rather irregular, e.~g. 
for trends due to the southern oscillation (El Nino) effect.  In both cases, 
the effects of the trends have to be distinguished from the intrinsic 
fluctuations of the system under consideration, and strong oscillatory
trends can lead to a false determination of the correlation behavior.  
In particular, it is of great interest to know, how strong the oscillatory 
trends may become until they start to disturb the correlation analysis.  

In order to investigate the effect of oscillatory trends on the results of
the DFA, we have constructed artificial records of long-range correlated 
random numbers with superimposed oscillatory trends of different frequency 
and strength.  The results for the modified DFA fluctuation function 
$F^{(n)}_{\rm mod}(s)$ are shown in Fig.~\ref{res6}.  The figure shows 
that the DFA is more sensitive to slowly varying trends, while quickly 
oscillating trends disturb the scaling behavior of the results much less.  
Thus, for example, the seasonal trends in temperature records have to be 
removed prior to the analysis (e.~g. by subtracting a daily mean temperature), 
while the modulating effect of the breathing does not significantly disturb 
the analysis of heartbeat records.

\subsection*{Acknowledgments}
This work has been supported by the Deutsche Forschungsgemeinschaft and the 
project "KLIWA" of the Bayerische Landesamt f\"ur Wasserwirtschaft.


\begin{thebibliography}{99}

\bibitem{henio} Present address: Dipartimento de Fisica, 
Universidade Federal do Rio Grande de Norte, 59072-970 Natal, Brazil.

\bibitem{peng94} C.-K. Peng, S.V. Buldyrev, S. Havlin, M. Simons,
 H.E. Stanley, A.L. Goldberger, Phys. Rev. E {\bf 49} (1994) 1685.

\bibitem{buldy95} S.V. Buldyrev, A.L. Goldberger, S. Havlin, R.N. Mantegna,
M.E. Matsa, C.-K. Peng, M. Simons, H.E. Stanley,
 Phys. Rev. E {\bf 51} (1995) 5084.

\bibitem{dna}  C.-K. Peng, S.V. Buldyrev, A.L. Goldberger, R.N. Mantegna, 
 M. Simons, H.E. Stanley, Physica A {\bf 221} (1995) 180.

\bibitem{dna1}  S.V. Buldyrev, N.V. Dokholyan, A.L. Goldberger, S. Havlin, 
 C.-K. Peng, H.E. Stanley, G.M. Viswanathan, Physica A {\bf 249} (1998) 430.

\bibitem{herz} C.-K. Peng, J. Mietus, J.M. Hausdorff, S. Havlin,
 H.E. Stanley, A.L. Goldberger, Phys. Rev. Lett. {\bf 70} (1993) 1343. 

\bibitem{peng95} C.-K. Peng, S. Havlin, H.E. Stanley, A.L. Goldberger,
 Chaos {\bf 5} (1995) 82.

\bibitem{herz1} K.K.L. Ho, G.B. Moody, C.-K. Peng, J.E. Mietus, M.G. Larson, 
 D. Levy, A.L. Goldberger, Circulation {\bf 96} (1997) 842.

\bibitem{viswan97} G.M. Viswanatha, C.-K. Peng, H.E. Stanley, A.L. Goldberger, 
 Phys. Rev. E {\bf 55} (1997) 845.

\bibitem{peng98} C.K. Peng, J.M. Hausdorff, S. Havlin, J.E. Mietus, 
 H.E. Stanley, A.L. Goldberger, Physica A {\bf 249} (1998) 491. 

\bibitem{herz2} Y. Ashkenazy, M. Lewkowicz, J. Levitan, S. Havlin, K. Saermark, 
 H. Moelgaard, P.E.B. Thomsen, Fractals {\bf 7} (1999) 85.

\bibitem{herz3} H.E. Stanley, L.A.N. Amaral, A.L. Goldberger, S. Havlin, 
 P.Ch. Ivanov, C.-K. Peng, Physica A {\bf 270}, 309 (1999).

\bibitem{herz4} P.-A. Absil, R. Sepulchre, A. Bilge, P. Gerard, 
 Physica A {\bf 272} (1999) 235.

\bibitem{herz5} P.C. Ivanov, A. Bunde, L.A.N. Amaral, S. Havlin, J. Fritsch-Yelle, 
 R.M. Baevsky, H.E. Stanley, A.L. Goldberger, Europhys. Lett. {\bf 48} (1999) 594.  

\bibitem{herz6} A. Bunde, S. Havlin, J.W. Kantelhardt, T. Penzel, J.-H. Peter, 
 K. Voigt, Phys. Rev. Lett. {\bf 85} (2000) 3736.

\bibitem{neuron} S. Blesic, S. Milosevic, D. Stratimirovic, M. Ljubisavljevic, 
 Physica A {\bf 268} (1999) 275.

\bibitem{neuron1} S. Bahar, J.W. Kantelhardt, A. Neiman, H.H.A. Rego, D.F. Russell, 
 L. Wilkens, A. Bunde, F. Moss (submitted, 2000).

\bibitem{gait} J.M. Hausdorff, S.L. Mitchell, R. Firtion, C.-K. Peng, 
 M.E. Cudkowicz, J.Y. Wei, A.L. Goldberger, J. Appl. Physiology {\bf 82} (1997) 262.

\bibitem{koscielny98} E. Koscielny-Bunde, H.E. Roman, A. Bunde, S. Havlin and 
 H.-J. Schellnhuber, Phil. Mag. B {\bf 77} (1998) 1331.

\bibitem{koscielny98b} E. Koscielny-Bunde, A. Bunde, S. Havlin, H.E. Roman, 
 Y. Goldreich, and H.-J. Schellnhuber, Phys. Rev. Lett. {\bf 81} (1998) 729.

\bibitem{talkner00} P. Talkner, R.O. Weber, Phys. Rev. E {\bf 62} (2000) 150.

\bibitem{cloud} K. Ivanova, M. Ausloos, Physica A {\bf 274} (1999) 349.

\bibitem{cloud1} K. Ivanova, M. Ausloos, E.E. Clothiaux, T.P. Ackerman, 
 Europhys. Lett. {\bf 52} (2000) 40.

\bibitem{economics} Y.H. Liu, P. Cizeau, M. Meyer, C.-K. Peng, H.E. Stanley,
 Physica A {\bf 245} (1997) 437.

\bibitem{economics1} P. Cizeau, Y.H. Liu, M. Meyer, C.-K. Peng, H.E. Stanley,
 Physica A {\bf 245} (1997) 441.

\bibitem{economics2} M. Ausloos, N. Vandewalle, P. Boveroux, A. Minguet, 
 K. Ivanova, Physica A {\bf 274} (1999) 229.

\bibitem{economics3} M. Ausloos, K. Ivanova, Physica A {\bf 286} (2000) 353.

\bibitem{fest} J.W. Kantelhardt, R. Berkovits, S. Havlin, A. Bunde, 
 Physica A {\bf 266} (1999) 461.

\bibitem{fest1} N. Vandewalle, M. Ausloos, M. Houssa, P.W. Mertens, M.M. Heyns,
 Appl. Phys. Lett. {\bf 74} (1999) 1579.

\bibitem{wtmm} J.F. Muzy, E. Bacry, A. Arneodo, 
 Phys. Rev. Lett. {\bf 67} (1991) 3515.

\bibitem{wtmm1} E. Bacry, J.F. Muzy, A. Arneodo, 
 J. Stat. Phys. {\bf 70} (1993) 635.

\bibitem{wtmm2} J.F. Muzy, E. Bacry, A. Arneodo, 
 Phys. Rev. E {\bf 47} (1993) 875.

\bibitem{randomwalk} M.F. Shlesinger, B.J. West, and J. Klafter,
 Phys. Rev. Lett. {\bf 58} (1987) 1100.

\bibitem{randomwalk1} D. Ben-Avraham and S. Havlin,  
 {\it Diffusion and Reactions in Fractals and Disordered System} 
 (Cambridge University Press, 2000).

\bibitem{hurst65} H.E. Hurst, R.P. Black, and Y.M. Simaika, 
{\em Long-term storage. An experimental study}  (Constable, London, 1965).

\bibitem{feder88} J. Feder, {\em Fractals} (Plenum Press, New York, 1988).

\bibitem{fano} U. Fano, Phys. Rev. {\bf 72} (1947) 26.

\bibitem{allan} J.A. Barmes, D.W. Allan, Proc. IEEE {\bf 54} (1996) 176.

\bibitem{taqqu95} M.S. Taqqu, V. Teverovsky, and W. Willinger, 
 Fractals {\bf 3} (1995) 785.

\bibitem{makse96} H.A. Makse, S. Havlin, M. Schwartz, and H.E. Stanley,
 Phys. Rev. E {\bf 53} (1996) 5445.

\bibitem{levywalk} P. Levy, {\it Calcul des probabilites} 
 (Gauthier-Villars, Paris, 1925), see also: R.N. Mantegna, H.E. Stanley, 
 {\it An Introduction to Econophysics} (Cambridge University Press, 2000).

\end{thebibliography}
\end{document}